\documentclass{aa}
\input{psfig.sty}
\def\simgt{\lower.5ex\hbox{$\;\buildrel>\over\sim\;$}}
\def\simlt{\lower.5ex\hbox{$\;\buildrel<\over\sim\;$}}
\begin{document}

   \thesaurus{03     (11.03.2; 
                      11.04.2; 
                      11.09.1 SBS~0335-052; 
                      11.19.3)} 
   \title{The near-infrared view of SBS~0335-052
\thanks{Based on data obtained at ESO-NTT in La Silla, Chile
and at UKIRT on Mauna Kea, Hawaii.}}


   \author{L. Vanzi\inst{1}, L. K. Hunt\inst{2}, 
T. X. Thuan\inst{3} and Y. I. Izotov\inst{4} 
          }

   \offprints{L. Vanzi}

   \institute{European Southern Observatory (ESO), 
              Alonso de Cordova 3107, Santiago - CHILE\\
              email: lvanzi@eso.org
          \and
              CAISMI-CNR, Largo E. Fermi 5, 50125 Firenze - ITALY
          \and
              Astronomy Department, University of Virginia, Charlottesville, VA 22903 - U.S.A.
          \and
              Main Astronomical Observatory, National Academy of Sciences
	      of Ukraine, Goloseevo, Kiev-127, 03680 - UKRAINE
             }

   \date{Received 29 May 2000; accepted 4 September 2000}

   \maketitle

   \begin{abstract}

We present near-infrared (NIR) spectra, deep $J$, $H$ and $K$ low-resolution images,
and high-resolution images in $J$ and $Ks$ of the low-metallicity blue dwarf galaxy SBS~0335-052. 
These new NIR data, together 
with optical and mid-infrared data from the literature, are used to constrain 
the properties and the star formation history of the galaxy. 
The NIR spectral characteristics and broadband colors of this galaxy 
are those of an extremely young starburst. 
We find that the NIR emission is
dominated by star formation occuring in clusters younger than 5 Myr; no
optically hidden star formation is revealed in our data. We also find
that a warm-dust component is necessary to explain the $H-K$ color,
and is consistent with the mid-infrared spectral energy distribution.
Finally, we quantify the possible contribution from an evolved stellar population. 
\keywords{Blue Compact Dwarf Galaxies --
          Near-Infrared --
          SBS~0335-052 }
\end{abstract}

%

\section{Introduction}

SBS~0335-052 (Izotov et al. 1990) is a Blue Compact Dwarf (BCD) galaxy
that, at an abundance of $Z_{\odot}/41$ (Melnick et al. 1992,
Izotov et al. 1997), is
the second lowest metallicity galaxy known after I~Zw~18. 
SBS~0335-052 hosts an exceptionally powerful episode of star formation,
and Thuan et al. (1997) proposed, on the basis of HST 
observations, that the present burst is possibly the first star-forming 
episode in the history of the galaxy. 
The same authors deduce that the star formation occurs in six Super-Star 
Clusters (SSCs) not older than 25 Myr and
located within a region smaller than 2\arcsec, or 520 pc at a distance of 
54.3 Mpc that we will assume in this paper
($v_{helio}\,=\,4060\,\pm\,12$\,km/s, see Izotov et al. 1997).
Though the presence of dust mixed with the star forming clusters is evident
from the HST images, the ISO observation by Thuan et al. 
(1999) of a very strong continuum at $10\mu$m was a surprise. This was
successfully modeled with a modified black body and a screen extinction 
of $A_V=21$, and interpreted as due to 
thermal emission by silicate dust (M\,$\simlt\,10^3\,M_\odot$)
warmed by embedded clusters of star formation 
completely invisible in the optical. 

Because of the lack of chemical enrichment, 
low-metallicity BCDs such as SBS~0335-052 are also hypothetical candidates for
young or primeval galaxies (PGs) undergoing their first burst of star formation.
While
the majority of BCDs show an underlying extended low-surface-brightness 
component with colors indicative of an intermediate-age or old stellar population
(Loose \& Thuan 1985; Kunth et al. 1988; Papaderos et al. 1996;
 Telles \& Terlevich 1997), there is still debate whether such a component
has been detected in the two most metal-deficient BCDs presently known,
I\,Zw\,18 (Z$_\odot$/50, Searle \& Sargent 1972) and SBS 0335--052. 
High-resolution HST optical images of I\,Zw\,18
(Hunter \& Thronson 1995) show colors of the underlying
diffuse component consistent with a sea of unresolved B or early A stars, with
no evidence for stars older than a few tens of Myr, although Aloisi et al. 
(1999) and \"Ostlin (2000) have claimed to detect a 1-4 Gyr stellar population 
in that BCD.
Similar observations of SBS\,0335-052 (Thuan et al. 1997) show colors much
bluer than those expected for such an intermediate-age stellar population.
The optical evidence is only suggestive, however, since a large fraction of
the extended emission in SBS\,0335-052 is of gaseous origin (Thuan et al. 1997).

\begin{figure*}
\psfig{figure=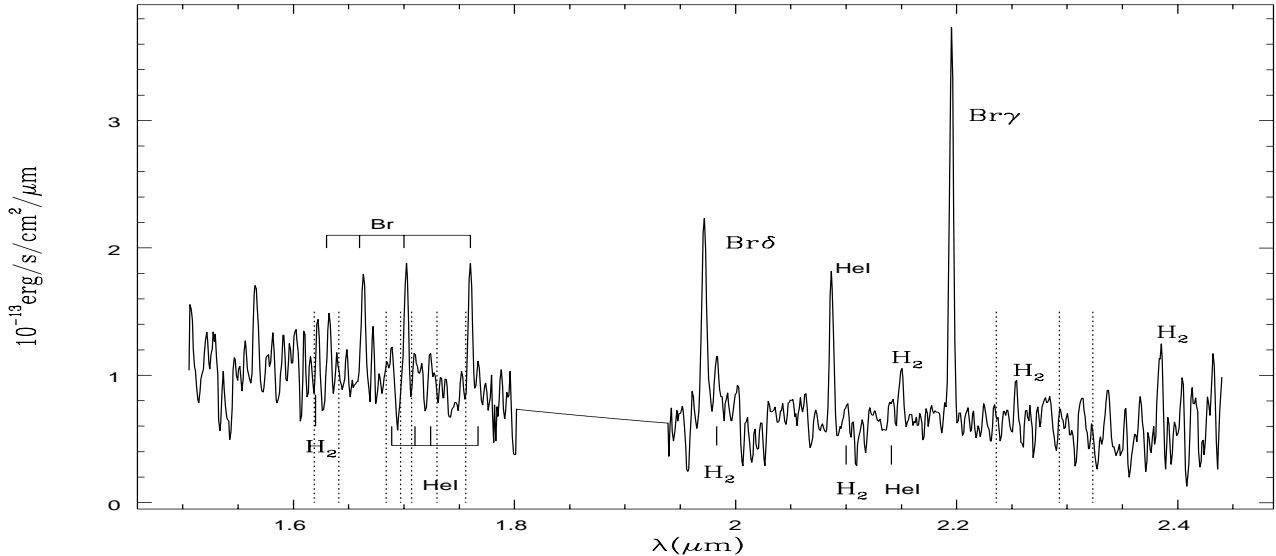,width=18cm,height=8cm,angle=-90}
\caption{Low-resolution near-infrared spectrum of SBS~0335-052. The vertical 
dotted lines show the positions of 
the strongest stellar absorption (CO) lines. The continuum slope has
been corrected according to the photometric measurements.}
\label{lowres}
\end{figure*}

Near-infrared (NIR) colors can help resolve the age question, since they are
extremely effective indicators of stellar population age, {\it given the
metallicity}.
While NIR colors of {\it evolved} stellar populations vary little
(Griersmith et al. 1982; Frogel 1985; Giovanardi \& Hunt 1988, 1996),
recent evolutionary synthesis models (Leitherer et al. 1999 --
SB99; Kr\"uger et al. 1995) show that the NIR colors of {\it young} stellar
populations differ significantly from those of older ones.
Both models find that longward of 1\,$\mu$m, the nebular continuum
dominates the emission during the early phases (a few Myr) of the starburst.
Therefore NIR colors provide a unique and effective diagnostic, 
and when combined with optical colors, are good indicators
of stellar population age.

NIR spectroscopy has a three-fold importance for our analysis. 
First, it
enables the estimate of the nebular contribution to the emission in order to
decontaminate the broadband colors.
Second, it is an almost extinction-free probe of the starburst parameters in what 
are typically dusty environments.
Finally, it offers unique constraints on the
star-formation history of the galaxy. In particular the equivalent width of
$Br\gamma$ is possibly the best age indicator for young starbursts since it
measures the ratio between young blue massive stars and the evolved red stellar population.
The NIR CO absorption features are the strongest signature of cold evolved
stars, as are the He ricombination lines for hot massive stars, and [FeII]
for supernova remnants. Finally the infrared is the only spectral region where the
molecular gas can be detected in its warm phase through $H_2$ emission lines.

It is therefore of crucial importance to further investigate SBS~0335-052
in the near-infrared to study the stellar populations
in this extremely low-metallicity system, as well as investigate the putative
hidden star formation suggested by the large extinction derived from
the mid-infrared spectral energy distribution. 
With this aim we have obtained NIR spectra in the $H$ 
and $K$ bands, and broadband images in $J$, $H$, $K$, and $Ks$. 
We present here the results of our new observations. 
The SBS\,0335-052 system consists of two widely separated (22\,kpc in the
East-West direction) star-forming components within a large 
64\,$\times$\,21\,kpc HI cloud (Pustilnik et al. 2000).
We denote the eastern component by SBS\,0335-052, and the western one 
(Lipovetsky et al. 1999) by SBS\,0335-052W.

The present paper is organized as follows. In Sect.2 we present our
data along with a description of the observations and data reduction. Sect.3
is a description of the NIR spectrum and of
the properties of the galaxy as can be derived from the analysis of the 
spectrum. The same is done in Sect.4 for the near-infrared images.
The characteristics of the galaxy derived from the observations are combined
with data from the literature, and then used to constrain a starburst model in
Sect.5. In Sect.6 we summarize the results of our work.

\section{Observations and data reduction}

Our data were acquired at ESO, La Silla, with the NTT and SOFI, and at UKIRT,
Mauna Kea, with IRCAM3.
The former camera$+$spectrometer is based on a 1024$\times$1024 Hawaii HgCdTe
array, and the latter camera on a 256$\times$256 NICMOS3.

\subsection{NIR spectra}

Low-resolution (R=600) spectra of SBS~0335-052 covering the $H$ and $K$ infrared 
bands were obtained with SOFI at the NTT in two different occasions: first
in August 1998 with 45m of integration, and later 
in December 1998 with similar conditions and the same amount of integration time.
In both cases we used a slit 1\arcsec\ wide (5~arcmin length)
oriented along the "major axis"
of the galaxy $PA\approx 145$. Since the source is much smaller than the
length of the slit, the nod-on-slit technique was used to acquire the sky
spectrum. The data reduction followed the standard process of flat-fielding
and sky-subtraction. A lamp spectrum was used to correct the 2D spectra
for distortions along the slit, and to have an accurate
wavelength calibration. 1D spectra were extracted with an aperture of 1\farcs5
centered on the brightest peak.

Particular attention was given to achieving a good estimate of the 
continuum slope. The 1D spectra were divided by the spectrum 
of a solar-type star observed under the same conditions to correct for the 
telluric absorption features. Then the solar spectrum was used to 
reestablish the original slope as described by Maiolino et al. (1996).
The two spectra have been averaged with equal weights.
The spectrum was flux calibrated extracting the photometry from the 
corresponding apertures in our H and K images. This also lead to the 
adjustment of the continuum that was otherwise too blue\footnote{This is not 
surprising since it is well known that the variability of the atmospheric
transmission makes difficult a reliable determination of the infrared continuum.}.
Comparison of the spectrum with the photometric values was always done by
convolving the spectrum with the filter bandpass measured in the laboratory. 
The 1D combined spectrum flux and wavelength calibrated is shown in Fig.
\ref{lowres}.

\begin{figure}
\psfig{figure=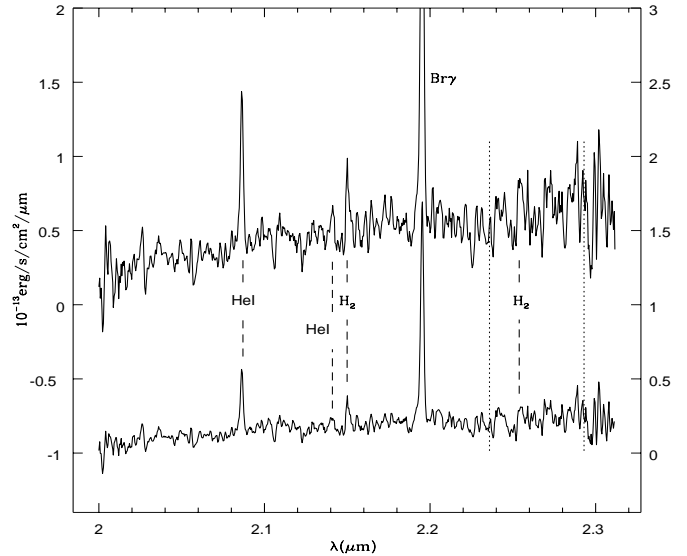,width=9cm,height=8cm,angle=0}
\caption{Medium-resolution near-infrared spectrum of SBS~0335-052. The scale on
the left refers to the top spectrum (aperture A), the one on the right to the
bottom spectrum (aperture B). The dotted lines mark the positions of stellar
absorption features. The slope of the continuum has not been corrected.}
\label{higres}
\end{figure}

A medium-resolution (R=1400) spectrum in the $Ks$ band was observed with SOFI
in October 1999 under exceptional seeing conditions (0\farcs5 average of the night);
the total integration time was in this case 1h. We used the same slit and
same PA as in the observations described above, and the same procedure 
was used for the data reduction. At this spatial resolution, two peaks of 
emission separated by about 1\farcs5 were clearly resolved in the 2D spectrum. 
Two 1D spectra centered on the two peaks were extracted with apertures of
0\farcs8 (south-east) and 0\farcs6 (north-west). We will refer to
these spectra as A and B respectively in the rest of the paper. 
Since the observed band
is well transmitted by the atmosphere no correction or readjustment of 
the slope was applied. The spectra were flux calibrated extracting the
photometry from the corresponding apertures in the 0\farcs073$K_S$ image 
(see next section). The flux- and wavelength-calibrated spectra
are shown in Fig.\ref{higres}.
All the spectra have been smoothed to lower resolution to reduce the noise.

Including the errors on the photometry, centering and size of the apertures
the flux calibration of the spectra can be estimated to be accurate within 10\%. 

\subsection{NIR images}

We acquired $J$ (1.2\,$\mu$m ), $H$ (1.6\,$\mu$m ), and $K$ (2.2 \,$\mu$m)
images of the nominal central (E) and western (W) component
of SBS\,0335-052 with the 3.8-m United Kingdom Infrared Telescope
(UKIRT\footnote{The United Kingdom Infrared Telescope is operated by the
Joint Astronomy Centre on behalf of the U.K. Particle Physics
and Astronomy Research Council.})
equipped with IRCAM3.
The plate scale is 0\farcs28, with a total field-of-view of 72\arcsec;
the E and W components are separated by roughly 84\arcsec, too large to be
accommodated by the IRCAM3 field-of-view, so
we performed separate sets of observations for the two components.
The source was placed in several different positions on the array, taking
care not to compromise the flat field by having overlap in the source
positions.
At the beginning of each observing sequence, 
dark exposures were acquired with the same parameters as the subsequent
science frames.
Total on-source integration times 
were 55\,min and 48\,min in $K$ for the E and W components, respectively;
30 and 34\,min in $H$; and 19 and 27\,min in $J$.
Individual frames were dark-subtracted and flat-fielded with a subset of
the remaining frames in the sequence, after editing them for stars (to
avoid "holes" in the reduced frames).
The reduced frames were then registered and averaged;
observations obtained on different nights were combined with a similar
algorithm, but with the flux scale given by the relative photometric
calibration, and weighted by the total integration time.

Photometric calibration was performed by observing standard stars,
every few hours, from the UKIRT Faint Standard List.
As for the source observations,
each standard was measured in several different positions on the array.
We corrected for atmospheric extinction by fitting simultaneously,
for each band, an extinction coefficient common to all nights
and a ``zero point'' that varied from night to night.
The resulting extinction coefficients are 0.049, 0, and 0.046 mag/airmass
in $J$, $H$, and $K$, respectively.
Formal photometric accuracy, as judged by the nightly dispersion of
the standard stars is 0.025\,mag, or better, in all three bands.
We have corrected all photometry for Galactic extinction ($A_B\,=\,0.09$~mag)
using the extinction law of Cardelli et al. (1989). 

\begin{figure}
\psfig{figure=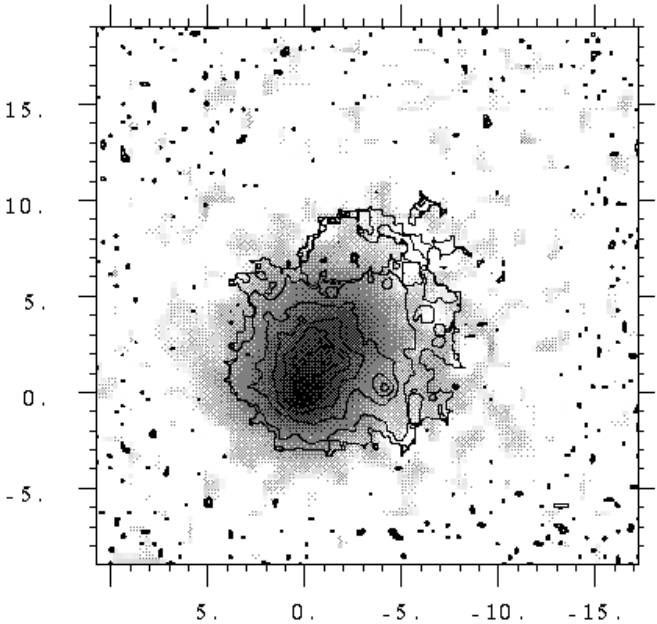,width=9cm}
\caption{The contours of 
the HST $V$-band image taken from Thuan et al. (1997), superimposed
on the $K$-band image of SBS~0335-052 obtained at UKIRT. 
The contours, shown in black, range from 16 to 24 $V$ mag/arcsec$^2$,
and the grey scale of the underlying image ranges from 16 to 21.5 $K$ 
mag/arcsec$^2$. The tickmark labels correspond to arcsec in the N and E 
directions.
\label{kimg_uk}}
\end{figure}

The final UKIRT deep $K$-band image is shown in
Fig.\ref{kimg_uk}, with the contours of the HST $V$-band image from Thuan et al. (1997)
superimposed. 
As measured from the combined frames,
seeing was 1\farcs1 in $J$, 0\farcs81 in $H$, and 0\farcs75 in $K$.
The combined images have a 1$\sigma$ sensitivity (per pixel) of
22.4 mag/arcsec$^2$ in $J$,
21.4 mag/arcsec$^2$ in $H$, and
21.4 mag/arcsec$^2$ in $K$.

\begin{figure}
\psfig{figure=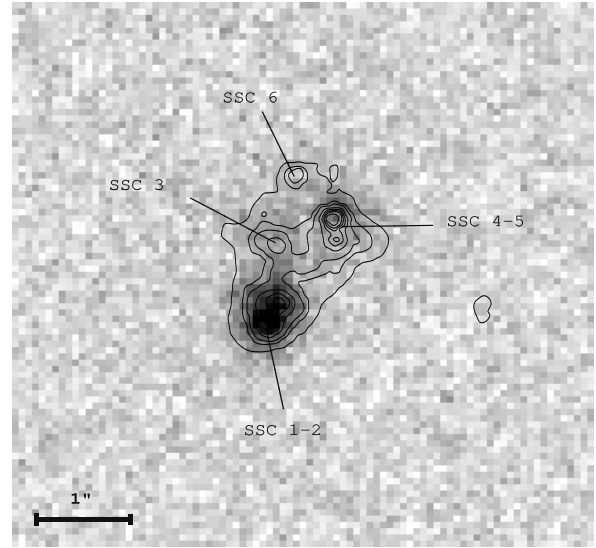,width=8.2cm}
\caption{The high-resolution (0\farcs073 /pixel) 
$Ks$-band image of SBS~0335-052 obtained with SOFI at NTT 
with the SSCs denoted by Thuan et al. (1997) labeled.
The HST $V$ image is superimposed on the SOFI image, with contours
running from 17 to 20 $V$ mag/arcsec$^2$, N up, E left.}
\label{kimg_hr}
\end{figure}

Three images were also acquired in October 1999 with SOFI at the NTT. 
A $J$ image
with a scale of 0.146 arcsec/pix and a total integration of 300 sec, a $Ks$
image with the same scale and integration time and a $Ks$ image with a scale
of 0\farcs073 /pixel and integration time of 400 sec. 
The ESO $K_s$ filter has a central wavelength $\lambda_c$ of 2.162\,$\mu$m, and
a FWHM $\Delta\lambda$ of 0.275\,$\mu$m
(c.f., UKIRT $K$ has $\lambda_c\,=\,2.21\,\mu$m, and FWHM$\,=\,0.37\mu$m).
The jitter-on-source
technique was always used. All the images were
reduced following the standard procedures of flat-field, sky-subtraction and
realignment. 
Photometric calibration was achieved through large-aperture photometry of
the UKIRT images.
The seeing measured on the combined images is 0\farcs49 ($J$),
0\farcs41 ($Ks$) and 0\farcs33 (0\farcs073 $Ks$)!
Hereafter, these SOFI images will be called high-resolution ($Ks$ only, 
0\farcs073/pix),
medium-resolution ($J$ and $Ks$, 0.146 arcsec/pix), and the UKIRT images
low-resolution ($J$, $H$, $K$ 0.28 arcsec/pix).
Fig.\ref{kimg_hr} shows the high-resolution $Ks$ SOFI image, together with the SSC
nomenclature of Thuan et al. (1997).
The figure shows, at a significantly higher resolution and better seeing than
Fig. \ref{kimg_uk}, the same central structure as seen in the visible and 
in the deeper UKIRT image.
The lack of extended structure in the high-resolution image is due to its
lower sensitivity.

\begin{table}
\caption[]{Recombination lines in SBS~0335-052 from the low-resolution 
spectrum, $\lambda$ in $\mu$m and fluxes measured in
10$^{-16}$\,erg/s/cm$^2$}
\label{h}
\begin{tabular}{ccc}
\hline
  Line   & $\lambda_{rest}$ & Flux \\
\hline
 Br13 & 1.611  & 1.9 $\pm$ 1.0 \\ 
 Br12 & 1.641  & 3.7 $\pm$ 1.0 \\ 
 Br11 & 1.681  & 4.4 $\pm$ 0.8 \\ 
 HeI$3^3P^0-4^3D$  & 1.700 & 1.0 $\pm$ 0.8 \\
 Br10 & 1.737  & 4.7 $\pm$ 0.8 \\ 
 Br$\delta$    & 1.945 & 9.0 $\pm$ 0.7 \\ 
 HeI$2^1S-2^1P^0$  & 2.058 & 5.3 $\pm$ 0.5 \\
 HeI$3^3P^0-4^3S$  & 2.113 & 1.2 $\pm$ 0.5 \\
 +$3^1P^0-4^1S$  &       &               \\
 Br$\gamma$    & 2.166 & 14.8 $\pm$ 0.8\\ 
\hline
\end{tabular}
\end{table}

\begin{table}
\caption[]{$H_2$ lines in SBS~0335-052 from the low-resolution spectrum,
$\lambda$ in $\mu$m and fluxes measured in 
10$^{-16}$\,erg/s/cm$^2$}
\label{h2}
\begin{tabular}{ccc}
\hline
  Line   & $\lambda_{rest}$ & Flux \\
\hline
 $H_2$(6,4)Q(1) & 1.601 & 1.5 $\pm$ 1.0 \\ 
 $H_2$(1,0)S(3) & 1.957 & 2.8 $\pm$ 1.0 \\ 
 $H_2$(1,0)S(2) & 2.033 & 1.0 $\pm$ 1.0 \\
 $H_2$(2,1)S(3) & 2.073 & 1.0 $\pm$ 0.7 \\
 $H_2$(1,0)S(1) & 2.121 & 2.4 $\pm$ 0.7 \\
 $H_2$(1,0)S(0) & 2.223 & 1.4 $\pm$ 0.7 \\
 $H_2$(2,1)S(1) & 2.248 & 1.0 $\pm$ 1.0 \\
 $H_2$(2,1)S(0) & 2.355 & 2.3 $\pm$ 1.0 \\
\hline
\end{tabular}
\end{table}

\begin{table}
\caption[]{Emission lines detected in the medium-resolution spectrum,
apertures A and B: $\lambda$ in $\mu$m and fluxes measured in units of 
$10^{-16}$\,erg/s/cm$^2$}
\label{hr}
\begin{tabular}{cccc}
\hline
  Line   & $\lambda_{rest}$ & Flux (A) & Flux (B) \\
\hline
 HeI$2^1S-2^1P^0$  & 2.058 & 2.5 $\pm$ 0.4 & 1.0 $\pm$ 0.2 \\
 HeI$3^3P^0-4^3S$  & 2.113 & 0.6 $\pm$ 0.4 & 0.2 $\pm$ 0.2 \\
 +$3^1P^0-4^1S$    &       &               &     \\
 $H_2$(1,0)S(1)    & 2.121 & 1.2 $\pm$ 0.4 & 0.5 $\pm$ 0.2 \\
 Br$\gamma$        & 2.166 & 8.9 $\pm$ 0.4 & 3.6 $\pm$ 0.2 \\ 
 $H_2$(1,0)S(0)    & 2.223 & 1.1 $\pm$ 0.6 & 0.3 $\pm$ 0.3 \\
\hline
\end{tabular}
\end{table}

\section{The NIR spectrum of SBS~0335-052}
The low-resolution spectrum of Fig.\ref{lowres} is
centered on the brightest spot observable in the galaxy, which includes 
the SSCs 1, 2 and partially 3 of the HST image of Thuan et al. (1997). 
The medium-resolution spectra of Fig.\ref{higres} acquired under
better seeing conditions are centered, respectively, on the SSCs 1--2 
and 4--5 following the same notation.

We detect two classes of lines in our near-infrared spectra, mainly 
ricombination lines and molecular hydrogen lines. Identification and flux
of the lines detected are shown in Tables \ref{h}, \ref{h2} and \ref{hr}.
The flux emitted in each line has been measured by integrating over the line
profile; since the lines are not resolved, we did not attempt to fit gaussians.

\subsection{Hydrogen lines and extinction}
The Br$\gamma$ in emission with an equivalent width (EW) exceeding 200~\AA~stands
out as one of the most prominent ever observed in any extragalactic
object. 
We have measured the extinction with this line, using
the ratio H$\beta$/Br$\gamma$. The 
H$\beta$ flux has been measured from the optical spectrum of Izotov et al. (1997),
extracted with the same aperture ($1\arcsec \times 1\farcs5$).
The value obtained is $3.273~10^{-14}$ erg/s/cm$^2$. Assuming an intrinsic
value of H$\beta$/Br$\gamma$ = 42.73 as calculated by Hummer \& Storey (1987) for
case B and T= 20000 K and $n_e=100~cm^{-3}$ 
(Izotov et al. 1997), and using the extinction law of Rieke \& Lebofsky 
(1985) we calculate $A_V = 0.73$.
This value is slightly higher than $A_V = 0.55$ 
measured in the optical (Izotov et al. 1997), but much lower 
than the ($A_V\,\sim\,21$)
estimate of Thuan et al. (1999) from their mid-infrared observations.
Other Brackett lines, although detected in the spectrum, do not have 
accurate enough flux measurements to give reliable estimates of the 
extinction on a relatively short wavelength base.

To compare our value with the optical one, we have extracted the H$\alpha$ flux 
in the same aperture from the spectrum of Izotov et al. (1997) to obtain
a flux of $1.04~10^{-13}$ erg/s/cm$^2$. With this value and using an
intrinsic ratio of 
H$\alpha$/H$\beta$ = 2.75 (Hummer \& Storey 1987), we calculate $A_V = 0.48$.
These values give E(B-V)(H$\beta$/Br$\gamma$)= 0.15 and E(B-V)
(H$\alpha$/H$\beta$) = 0.24 that are too similar to be able to distinguish the dust
distributions. 
Specifically, they are consistent with a homogeneous
foreground screen, a clumpy screen, or a homogeneous mixture of dust and
gas (Calzetti et al. 1996). If we repeat the same exercise using the
global fluxes from the galaxy integrated along the slit we derive the following
values: $2.20~10^{-13}$, $7.20~10^{-14}$ and $2.31~10^{-15}$ erg/s/cm$^2$ respectively
for H$\alpha$, H$\beta$ and Br$\gamma$ fluxes, and calculate
$A_V = 0.33-0.34$. We have therefore the possibility of a slightly higher
extinction toward the center of the galaxy, with the near-infrared penetrating
deeper into the core of the star formation. 
From the comparison of the extinction measured in the optical,
near- and mid-infrared we can conclude that even at $2\mu$m we are not
yet seeing the hidden star formation hypothesized by Thuan et al. (1999). 

\subsection{Molecular hydrogen lines}
Several emission lines of molecular hydrogen are detected in our 
low-resolution spectrum while only the (1,0)S(1) and (1,0)S(0) lines are clearly
visible in the medium-resolution spectrum.
We have compared our line ratios with the models summarized by Engelbracht et al.
(1998) in their Table 8 and derived from Black \& van Dishoeck (1987). 
With two exceptions, all the lines observed in the $K$ band are consistent, 
within the errors, with both thermal and fluorescent excitation. 
The (2,1)S(3) at $2.073\mu$m line should not be visible in our spectrum 
for thermal excitation and the (2,1)S(1) line $2.248\mu$m should be
relatively bright  for fluorescent excitation, while in our spectra it
is barely detected. Longward of rest wavelength 2.23 $\mu$m, the 
noise of the low-resolution spectrum increases due to the thermal background,
and this could explain the low detection of the (2,1)S(1) transition and thus
favor fluorescent excitation.

Evidence to support fluorescence comes from the $H$ part of
the spectrum where the transition (6,4)Q(1) is detected, though at
a low level, at rest wavelength 1.601 $\mu$m. 
The detection of this line is a bit doubtful, however, since it is in a quite 
noisy part of the spectrum.
Nevertheless, if confirmed it would be an unquestionable signature 
of fluorescent excitation. 
The non-detection of the transitions (5,3)O(3) and (6,4)O(3), that should be 
comparable in flux to (6,4)Q(1), is not surprising since at this resolution
they are respectively blended with Br13 and Br10. The quality of the spectrum
is not sufficient to attempt any deblending.
On the contrary, the non-detection of the (1,0)S(2)
transition in the high-resolution spectrum is surprising since the line should be
relatively bright for both fluorescent and thermal excitation.
We can therefore draw no firm conclusions
in either direction based on the 
analysis of the $H_2$ line ratios.

Whether fluorescent or thermal, the $H_2$ excitation is very likely to be
related to the strong UV field that dominates SBS~0335-052 as clearly shown by the 
outstanding Br$\gamma$. 
If we assume that UV exitation dominates, we can use the ratio
(1,0)S(1)/Br$\gamma$ as a diagnostic. Both lines in fact would have the 
same origin,
mainly UV radiation by massive stars, and their ratio is defined by the
efficiency of the $H_2$ excitation. Puxley et al. (1990) have shown that
this efficiency is a function of the gas density and of the UV flux strength.
The latter parameter is tightly bound to the IMF characteristics and to
the geometry of the emitting region. In particular they show how high 
values of (1,0)S(1)/Br$\gamma$,
typically 0.4 - 0.9, are easily produced by an ensemble of hot stars embedded
in a large molecular cloud (geometry A), while low values, below 0.4, are 
most likely produced by large ``nude" clusters of stars since the 
ratio of the $H_2$ emitting surface and the HII volume is minimized 
(geometry B). 
Vanzi \& Rieke (1997) have measured (1,0)S(1)/Br$\gamma$ typically below 0.1
in their sample of BCDs which suggests that star formation occurs in clusters in
these galaxies. This is perfectly consistent with what is observed for
instance in NGC~5253, II~Zw~40 (Aitken et al. 1982) and SBS~0335-052 itself.

Here, we can compare the measured values of (1,0)S(1)/Br$\gamma$ 
(=\,0.16, 0.13 and 0.14, respectively for the low-resolution, 
medium-resolution A and medium-resolution B) with the models of Puxley 
et al. (1990). From the H$\beta$
flux corrected for 0.55 magnitudes of visual extinction,
we have a ionizing flux Log(UV)=52.6 (case B). 
For $n_e=10^2~cm^{-3}$ the observed ratios are well reproduced by geometry A 
but a slightly higher density (that is a very likely possibility since Izotov 
et al. 1997 quote 390 $cm^{-3}$)  would cause the ratio to jump to unacceptably 
high values. A density higher than $10^2~cm^{-3}$ or a higher UV 
flux would be necessary to reproduce the observed value in geometry B.
Given the density measured by Izotov et al. (1997) the latter is therefore 
the most likely picture. 

It is remarkable that the emission in the (1,0)S(1) line is so
strong when compared to other low-metallicity BCD galaxies as those 
observed by Vanzi et al. (1996) and Vanzi \& Rieke (1997).
Moreover, the 
(1,0)S(1)/Br$\gamma$ is amongst the highest observed for this class of
galaxies. This can be possibly justified with a high molecular gas
($H_2$) content in
SBS~0335-052 as the dust abundance suggested by Thuan et al. (1999) may
support. 
The role of dust in the $H_2$ emission is, in any case, far from being
understood and any statement in that respect must be considered, at the 
moment, simply speculative. 
The $H_2$ is probably very clumpy as FUSE observations of I\,Zw\,18
(Vidal-Madjar et al. 2000) have shown that there is very little diffuse $H_2$
($<\,$30\,$M_\odot$).

\subsection{Helium lines}
We detect
the $4^3D-3^3P^0$ transition at rest wavelength 1.700 $\mu$m and the blend of
$4^1S-3^1P^0$ and $4^3S-3^3P^0$ at rest wavelength 2.113 $\mu$m.
The blend at 2.113 $\mu$m is also detected in the high resolution 
spectrum on both apertures.

It has been pointed out by Vanzi et al. (1996) and later by Engelbracht et al. 
(1998), how the 1.70 and 2.11 $\mu$m He I line can be used to constrain the presence of massive
stars in starbursting galaxies. In fact their ratio with the HI recombination
lines is proportional to the relative volumes of ionized He and H and
therefore to the temperature of the ionizing stars. Above a maximum value
of the effective star temperature (40000K, Shields 1993),
the two volumes become identical and the line ratio saturates. 

For a gas temperature of 20000 K and a singly ionized helium abundance of 0.078 
(Izotov et al. 1997) the saturated values calculated by Vanzi et al. (1996)
rescale as follow HeI\,1.7/Br10 = 0.29 and HeI\,2.11/Br$\gamma$ = 0.049.  
Both observed ratios are consistent with the saturated values which implies 
that stars with $T_*>40000K$, or $M>35 M_{\odot}$, are present in large
numbers in all the regions observed.

We do not attempt to interpret the HeI\,2.06/Br$\gamma$ ratio since this is
complicated by radiative transfer effects. We simply notice that the 
observed values 0.36, 0.26 and 0.24, respectively in the three spectra,
are consistent with the model of Shields (1993) with $T_*>45000~K$, 
$n=100~cm^{-3}$ and filling factor unity.

The positions of other HeI recombination lines are marked in the $H$ part
of Fig.\ref{lowres}. Peaks are detected at the correct wavelengths,
but though the correspondence might be tantalizing, these detections
are not convincing since they are at the 1 $\sigma$ level at best. Furthermore
all these lines should be much fainter according to the ratios
of Smits (1991). 
However it is known (Izotov et al. 1999) that optical He I lines are 
enhanced by fluorescence and collisional mechanisms and such mechanisms
could also be potential 
sources of the enhancement of He I lines also in the NIR. Benjamin et al. 
(1999) have shown that collisional enhancement is relatively small for the
strongest lines HeI\,1.70, 2.06 and 2.11 $\mu$m, but they do not 
discuss the effect on other lines. Unfortunately
the NIR spectrum is not deep enough to claim with certainty the detection of He I
lines other than those discussed above and to quantify the enhancement.

\subsection{Other spectral features}
Besides the recombination and molecular hydrogen lines, no other relevant
features are detected in our spectra. Some conclusions can be driven by
these non-detections.

The complete absence of the [FeII]1.64 line in the $H$ part of the spectrum
is clear evidence for a low number of supernovae in SBS~0335-052. The relation
between the emission of this line in starburst galaxies and the presence of
supernova remnants (SNRs) has been extensively proven and does not need to be 
discussed here (see for instance Moorwood \& Oliva 1988;
Greenhouse et al. 1991; Forbes \& Ward 1993; Vanzi \& Rieke 1997).
Given the low metallicity of SBS~0335-052 it is impossible to convert the
[FeII]1.64 upper limit into a reliable limit on the number of supernovae. 

The position of the main stellar absorption features are marked in Fig.
\ref{lowres} and \ref{higres} by dotted lines.
No clear stellar absorption bands are detected either in the $H$ or in the
$K$ spectrum. The depression of the continuum between 2 and 2.05 $\mu$m
corresponds to the main absorption band in $K$ and is surely due to the
earth's atmosphere. 
Stellar absorption bands in this spectral range are dominated by CO
and metal absorption lines in the atmosphere of cold evolved stars. Though the
S/N on the continuum is low, this proves at least that the contribution to
the continuum by evolved stars must be small.
Both the small number of SNRs and the lack of stellar absorption features
support the idea that SBS~0335-052 is dominated by a very young
population of massive stars, a point that will be better quantified in Sect.5. 

We also detect an emission line at restwavelength 1.545 $\mu$m, that
we have not been able to identify with any known line; it
is by far too bright to be identified as Br17 (no Brackett line is
above our detection limit beyond Br13) and is possibly spurious.

\begin{figure}
\psfig{figure=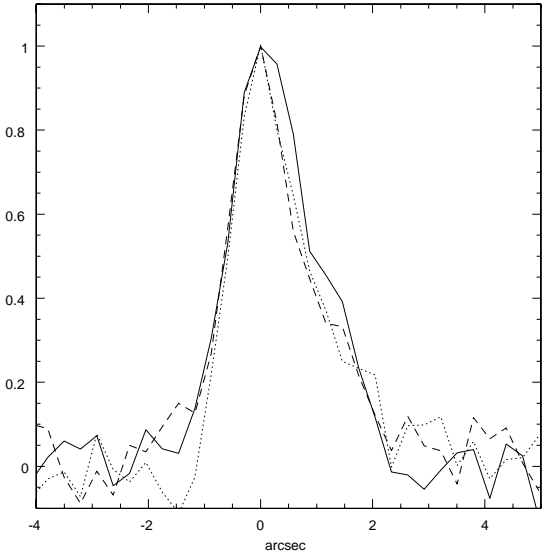,width=8cm,height=8cm}
\caption{Spatial profile along the slit of the K continuum (solid), 
Br$\gamma$ (dashed) and $H_2$ (dotted) in the low resolution spectrum of
SBS~0335-052.}
\label{lrprof}
\end{figure}

\begin{figure}
\psfig{figure=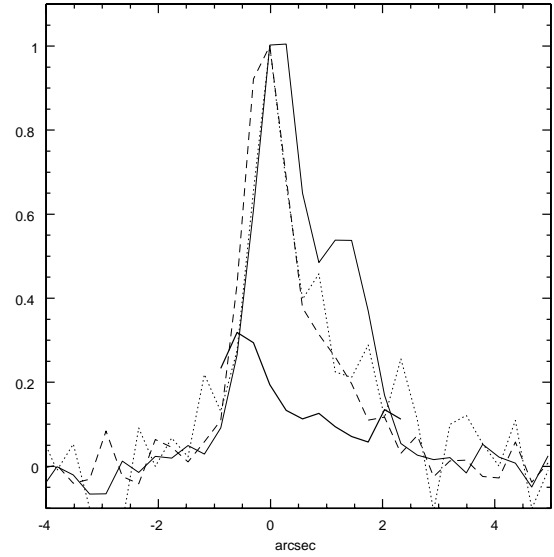,width=8cm,height=8cm}
\caption{Spatial profile along the slit in the medium resolution spectrum
of SBS~0335-052, lines as in the previous image plus Br$\gamma$/continuum 
as a solid thick line.}
\label{hrprof}
\end{figure}

\subsection{Spatial profiles of emission lines}
From our 2D spectra we have extracted the profiles corresponding to Br$\gamma$,
$H_2$2.12 and the continuum between these two emission lines. The results
are plotted in Fig. \ref{lrprof} for the low-resolution spectrum and
in Fig. \ref{hrprof} for the high-resolution spectrum. All profiles
are normalized to unity. Though the spectral resolution is 
different, the spatial scale is identical in the two observations.

In Fig. \ref{lrprof}, the seeing was about 1 arcsec. There is an evident 
difference between the continuum (solid line) and the emission lines (dotted and
dashed lines), with the continuum being more extended in the northwest direction.
This result only partially confirms the observation of Izotov et al. 
(1997) who find a shift of the optical continuum respect to the emission
lines of 200 pc in the northwest direction. Br$\gamma$
and $H_2$2.12 have an identical profile which supports the idea of a common
origin for the two lines, which must be mainly related to the star formation process.  

We can do a much better analysis on Fig. \ref{hrprof} since the spectrum
was taken under seeing conditions that were at least a factor of two better.  
In this case the difference is even more striking:
the continuum has a double-peaked shape. The two peaks can be
identified with the SSCs 1--2 and 4--5 of Thuan et al. (1997).
The emission lines, instead, have a single-peaked profile that is shifted by 0.23
arcsec, or 60 pc, in the southeast direction with respect to the main
continuum peak. This shift is at least a factor of three smaller than the one
measured by Izotov et al. (1997) in the optical and that has been recently 
confirmed by new high resolution optical observations at Keck (Izotov 2000, 
private communication).
The near-infrared continuum is therefore displaced respect to the optical.
We can interpret this result as due to extinction or to a different mechanism 
dominating the emission at different wavelengths. Since extinction is low
both in the optical and NIR,
the most likely explanation for this shift is the dominant gaseous
contribution to the 2\,$\mu$m emission (see Sect.4.3).

Again both emission lines show an identical profile that smoothly
declines with a wing toward northwest without showing a second peak.
The difference in shape between the continuum and the emission lines is
remarkable. 
We have directly compared our profiles with those of Izotov et al. (1997),
and after smoothing the
profiles of Fig.\ref{lrprof} to the lower spatial resolution, 
found a very good agreement between Br$\gamma$, $H_2$, H$\alpha$, H$\beta$
and $K$ continuum, while the shift in the optical continuum 
peaked 200~pc away from all the emission lines and the $K$ continuum
is striking.
We also attempted to study the H$\beta$/Br$\gamma$ ratio along the slit but
due to the low spatial resolution of the H$\beta$ profile this study did not 
add anything to the results presented in Sect.3.1. 

In Fig.\ref{hrprof} we also show the ratio Br$\gamma$/Continuum (thick
solid line) that has a main maximum southeast and a secondary peak between
the SSCs 1-2 and 4-5. This ratio tracks the star formation along
the slit.


\section{Near-infrared morphology and colors}

Fig.\ref{kimg_uk} and \ref{kimg_hr}
show that the NIR structure in the eastern (central)
component is very similar to that seen in the optical:
namely two peaks aligned along the southeast-northwest direction, 
separated by roughly 1\farcs5, and a northwest extended region.
This extension, in which filamentary structure is clearly 
evident in the optical, is smoother in the $K$ contours, but such smoothness
may be due to the limiting sensitivity of the $K$ image for 
extremely blue colors.
The peaks are the SSCs 1--2 and 4--5 in the notation of Thuan et al. (1997),
the separate components of which are not well resolved even in the
0\farcs07 $Ks$ image at 0\farcs3 seeing.
Also as in the optical, but slightly smaller,
there is a diffuse low-surface brightness region extending northwest of
the SSCs 4--5. 
Our UKIRT $K$ image is deep enough to sample typical galaxy colors of 
$B-K\,\sim\,3-3.5$,
so the $B-K$ color in the faint extension must be bluer than that.
Any structure in the western component, 84\arcsec\ away from the central (E) one,
is not easily discernible because
of its faintness and the low resolution of the deep UKIRT images.

\subsection{Integrated and SSC photometry }

From the deep UKIRT images
we derived total magnitudes of SBS~0335-052 and SBS~0335-052W which 
are given in Table \ref{tbl:phot_all}.
Typical formal errors on the photometry are 0.02--0.03 for the eastern
component, and 0.04--0.10 for the western one.
Inspection of the table shows that the NIR integrated colors
do not resemble an evolved stellar population,
which we would expect to have colors $J-H\,\sim\,0.7$
and $H-K\,\sim\,0.2$ (e.g., Frogel 1985).
Moreover, these $JHK$ colors are very unusual compared with those of other
BCDs (Thuan 1983). The two components have significantly different 
global $H-K$ colors with the eastern component 
being $\sim\,0.4$~mag redder.

We have also performed photometry on the high-resolution ($Ks$) images,
the medium-resolution images ($J$ and $Ks$), and
the low-resolution images ($J$, $H$, $K$) in
order to obtain photometry and colors of the SSCs 1+2 and 4+5 separately.
Although the two sets of SSCs are not well resolved with the UKIRT images,
we wanted to estimate the $J-H$ color, but
problems with seeing and the proximity of the SSCs
makes these magnitudes and colors uncertain.
In any case, it appears evident that the colors are anomalous,
but consistent with the global ones. 

\begin{table}
\caption[]{NIR Photometry of SBS\,0335-052\label{tbl:phot_all}}
\footnotesize{
\begin{tabular}{lcccc}
\hline
Component & $\phi^a$ &  $K_0^b$ & $(J-H)_0^b$ & $(H-K)_0^b$\\
\hline
Global (E)&  4 & 15.86 &  0.26 &  0.56\\
&  6 & 15.72 &  0.22 &  0.57\\
&  8 & 15.64 &  0.21 &  0.59\\
& 10 & 15.59 &  0.21 &  0.58\\
& 12 & 15.51 &  0.21 &  0.62\\
& 15 & 15.43 &  0.18 &  0.66\\
Outer (E)& -- & -- &  -0.06 &  0.44\\
\\
Global (W) &  4 & 18.72 &  0.19 &  0.27\\
&  6 & 18.27 &  0.30 &  0.21\\
&  8 & 17.95 &  0.31 &  0.19\\
& 10 & 17.75 &  0.37 &  0.17\\
Outer (W)& -- & -- &  0.36 &  0.29\\
\\
{\it {\bf Eastern:}}\\
NW extension & -- & -- & 0.06 & 0.58\\
\\
SSC 1+2 & 0.43$^c$ & 18.45 & -- \\
        & 0.88 & 17.55 & 0.78$^d$ \\
        & 1.17 & 17.29 & 0.69$^d$ \\
        & 1.4$^e$  & 16.97 & 0.31 & 0.68 \\
\\
SSC 4+5 & 0.43$^c$ & 19.32 & -- \\
        & 0.88 & 18.37 & 0.44$^d$ \\
        & 1.17 & 18.11 & 0.30$^d$ \\
        & 1.4$^e$ & 17.34 & 0.32 & 0.39\\
\\
Slit SE & 1$\times$0.8 & 17.24 & -- & -- \\
Slit NW & 1$\times$0.6 & 18.31 & -- & -- \\
\hline
\end{tabular} }

\noindent
$^a$\ Aperture diameter in arcsec.

\noindent
$^b$\ Corrected for Galactic extinction as described in text.

\noindent
$^c$\ SOFI high-resolution image at 0.072 arcsec/pixel.

\noindent
$^d$\ $(J-K)_0$, not $(J-H)_0$, taken from SOFI medium-resolution
images at 0.146 arcsec/pixel.

\noindent
$^e$\ UKIRT low-resolution images at 0.280 arcsec/pixel.
\end{table}

\subsection{Surface brightness profiles and outer colors }

\begin{figure*}
\hbox{\hspace{0cm}\psfig{figure=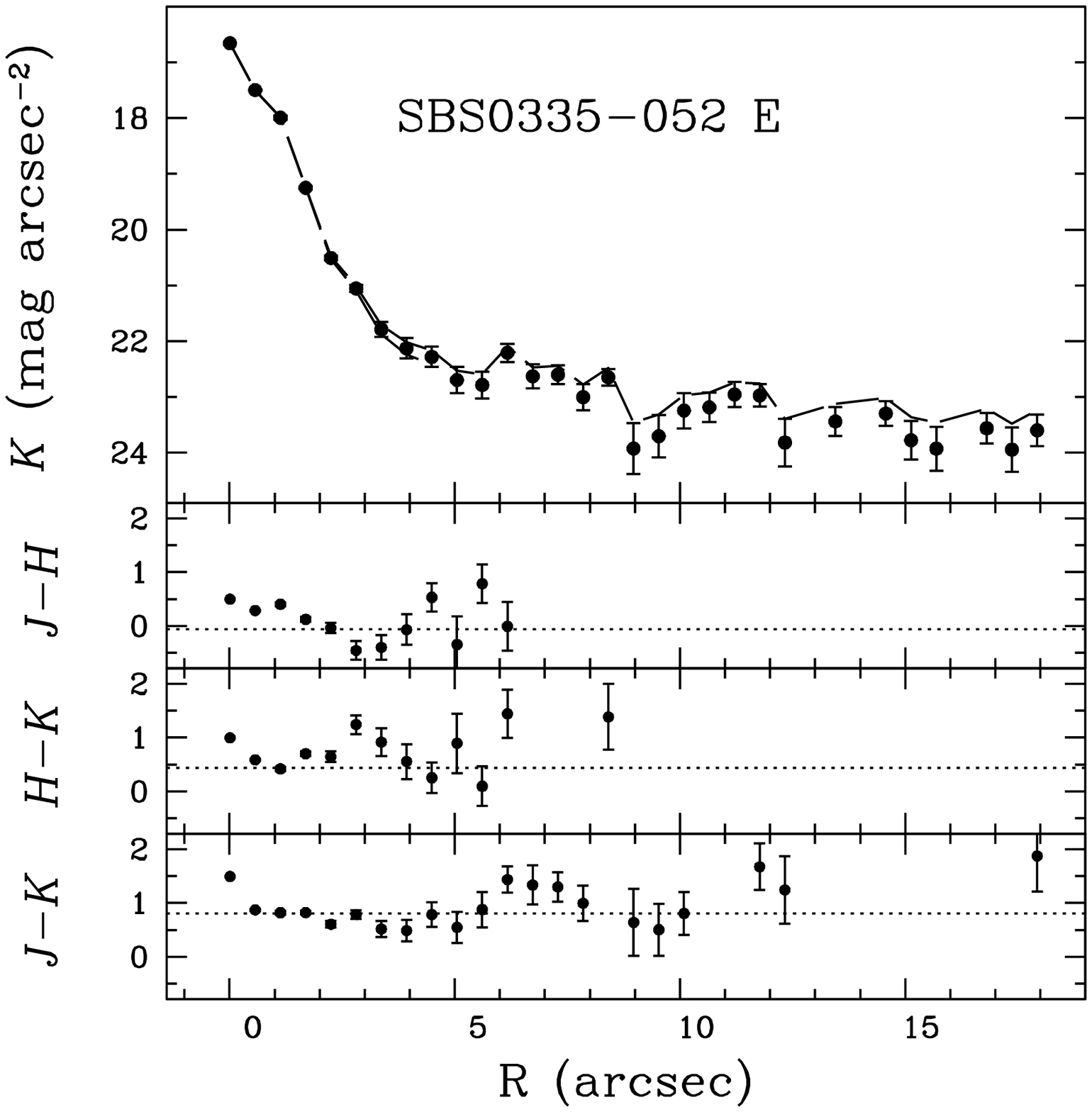,width=9.0cm}\hspace{0cm}
\psfig{figure=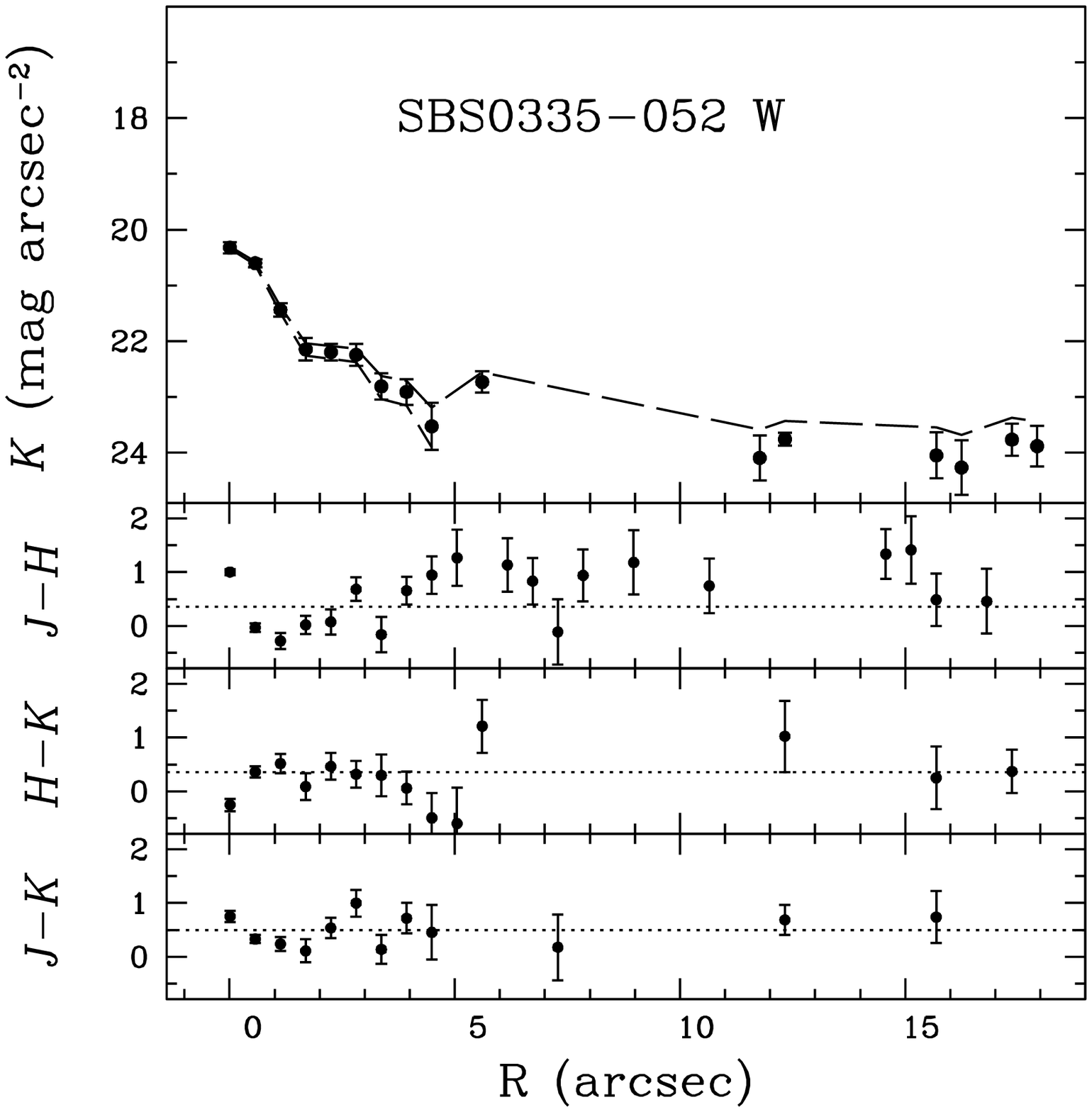,width=9.0cm}}
\caption{Radial surface brightness and color profiles of SBS~0335-052 
and SBS~0335-052W.
The upper panels show the $K$-band profiles, while the lower ones
show $J-H$, $H-K$, and $J-K$.
The dashed line in the upper panel shows the uncertainties of the sky
subtraction, and the dotted lines in the lower panels show the
mean outer colors as explained in the text.}
\label{profiles}
\end{figure*}

We have derived surface brightness profiles of both
SBS~0335-052 and SBS~0335-052W
by extracting circular annuli; these are essentially equivalent to
the derivative of the photometric growth curve.
We have used circular annuli instead of ellipse fitting because
the structure in the source is not amenable to ellipse fitting,
and we mainly wanted to analyze the profiles for their asymptotic
colors.
The circular profiles of SBS~0335-052 are centered on the surface brightness
peak, which corresponds roughly to SSC~1--2. 
These profiles, shown in Fig. \ref{profiles}, have been smoothed to two pixels
(0\farcs56) and extend to 23~mag/arcsec$^2$ in $K$.

The mean colors in the outer regions, also reported in Table \ref{tbl:phot_all}
as ``Outer",
are shown in the color panels as horizontal dotted lines;
they have been averaged over the low surface-brightness regions 
in both components ($\mu_K\,\ge\,20.5$ in the brighter E,
and $\mu_K\,\ge\,21.5$ in the fainter W).
Like the integrated colors, these mean outer colors of 
$J-H\,=\,-0.06\,\pm\,0.3$
and $H-K\,=\,0.44\,\pm\,0.5$ in the eastern component,
and
$J-H\,=\,0.36\,\pm\,0.4$ and $H-K\,=\,0.29\,\pm\,0.2$ in the 
western one are unusual, also relative to other BCDs, 
and are not indicative of an evolved stellar population.

To check the outer colors derived from the profiles, 
we have also performed photometry on the faint
region to the northwest in SBS~0335-052, beyond SSC~4--5.
These results are given in Table \ref{tbl:phot_all} as ``NW extension", 
and are, within the errors, consistent with the colors derived from
the radial profiles: namely $J-H\,\approx\,0$ and $H-K\,\approx\,0.4-0.6$.

\subsection{Gaseous contribution to the NIR emission}

It is well known that the nebular emission in young starbursts
is relatively strong in the near-infrared, and in some cases
may dominate the broadband flux
(Olofsson 1989; Leitherer \& Heckman 1995; Kr\"uger et al. 1995).
Hence, before comparing the NIR colors of SBS~0335-052 with evolutionary 
model predictions, it is useful to determine what fraction of the NIR emission
is due to gas, rather than stars.
To do this,
we have used the $Br\gamma$ flux\footnote{And assumed that the lines
are optically thin, which appears to be a reasonable assumption for the
NIR recombination lines in this source.}
in our two medium-resolution
spectra to estimate the continuum emission in those positions,
namely SSC~1--2 (A) and SSC~4--5 (B).
From the fluxes given in Table \ref{hr}, and
with the emission coefficients given by Joy \& Lester (1988)
for the physical conditions in SBS~0335-052 (20000K, He$^+$/H$^+$ = 0.078:
Izotov et al. 1997), we find a high percentage of gas emission
in both SSC groups:
47\% of the $Ks$ emission in the spectroscopic aperture centered on 
SSC~1--2 is nebular continuum, and 51\% in the aperture centered on 
SSC~4--5. Moreover, the $Br\gamma$ line alone contributes 6\% of the $Ks$
flux of SSC~1--2 and 7\% of SSC~4--5.
Therefore, the NIR colors of the SSCs are highly contaminated
by gas emission.


\subsection{Near-infrared/optical colors}
\begin{figure}
\psfig{figure=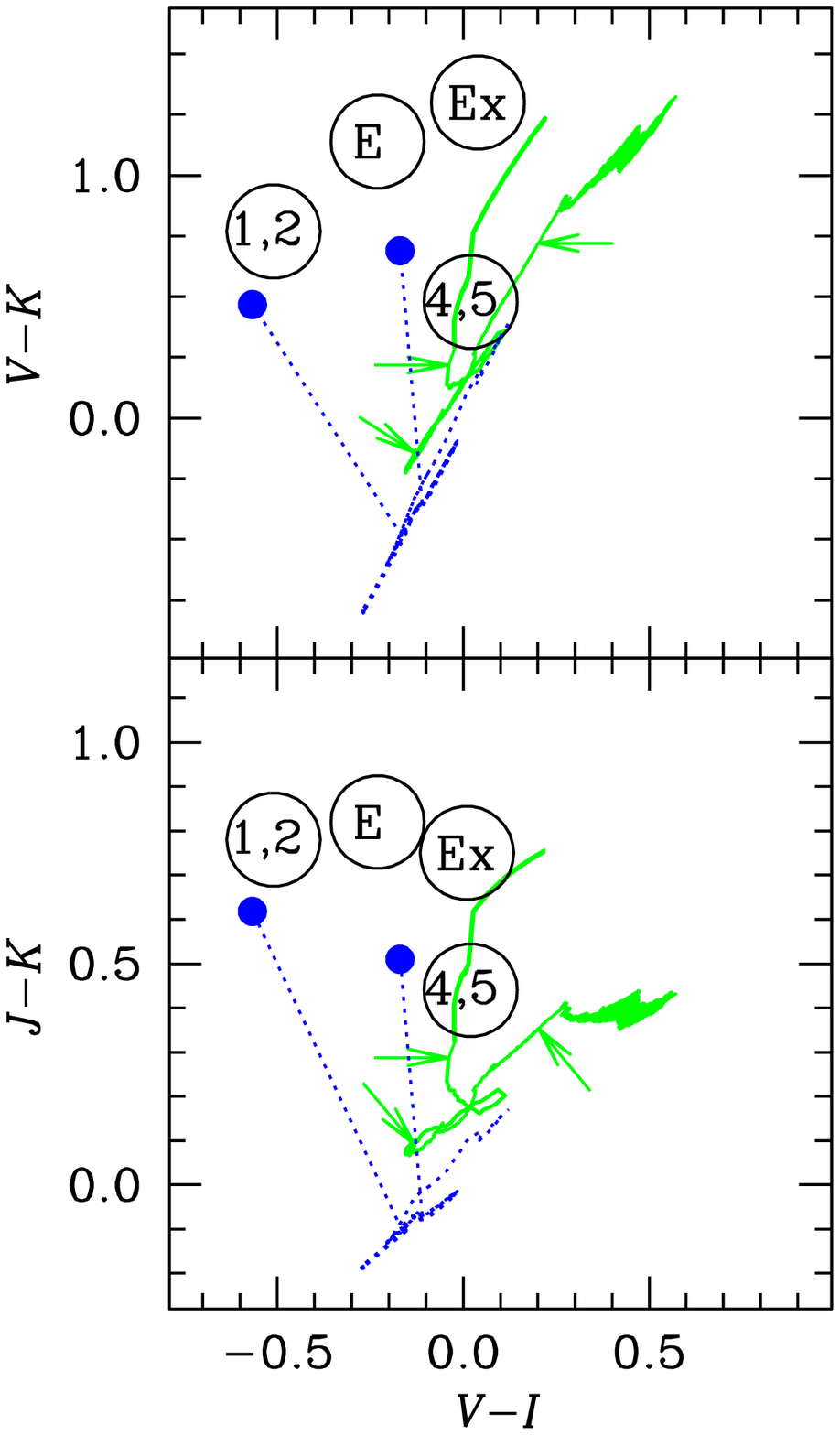,width=13cm,height=15cm,bbllx=5cm,bblly=6cm,bburx=22cm,bbury=25cm,angle=0}
\caption{Near-infrared/optical color-color diagrams: 
the upper panel shows $V-K$ vs. $V-I$, and the lower panel $J-K$ vs. $V-I$.
The colors are measured globally (``E"), 
for the SSCs 1+2, 4+5, and for the extended region (``Ex" in the figure). 
Also shown with solid lines are the evolutionary synthesis models for the 1/20 $Z_\odot$
instantaneous starbursts taken from SB99. 
The arrows denote transitions at 4~Myr, 10~Myr, and 25~Myr, with the
heavier lines showing the younger times.
Our models for the 1/50 $Z_\odot$ stellar population 
are shown as dotted lines; the lines connecting the stellar tracks
to the filled circles shows the color change necessary to accommodate
the observed gas contribution to a 3~Myr and 5~Myr (with extinction) stellar population.
}
\label{vmijvk}
\end{figure}

Fig.\ref{vmijvk} shows the color-color diagrams for $J-K$ (lower panel)
and $V-K$ (upper panel) versus $V-I$ (taken from Thuan et al. 1997).
Observed colors not corrected for gas emission are plotted. 
The labels inside the large open circles indicate the global colors
(``E''), the SSCs included in the measurement, and the extended region (``Ex").
Also shown (with solid lines) are the Leitherer et al. (1999) model predictions for a 
low-metallicity ($Z_{\odot}/20$) instantaneous starburst.
As explained in the figure caption, the arrows denote times of 4~Myr,
10~Myr, and 25~Myr after the burst, with the youngest times being shown by
the heaviest lines.
The $J-K\,\sim\,0.8$ of the outer regions (``Ex''), as measured
by the asymptotic colors in the radial $J-K$ profile 
is similar to that of SSC~1+2 and the global $J-K$;
it is also comparable to the $J-K\,\sim\,$0.7 of the region NW
of SSCs\,4--5 (``NW'') measured from the $J-K$ color image and reported
in Table \ref{tbl:phot_all}.

Even though the SB99 models take nebular emission into account, they do not
predict the colors we measure in this source.
Part of the reason for this may lie in its extremely low metallicity,
since the SB99 models are calculated for $Z\,=\,1/20\,Z_\odot$, not
1/40 $Z_\odot$ as in SBS~0335-052.
One consequence of such low metallicity that could cause deviations from the
models is the resulting higher-than-normal electron temperature which
changes the colors of the gas emission (to slightly bluer ones);
therefore, the correction implicit in the SB99 models would be inappropriate.
Another contribution to the discrepancy, probably the most important,
is the extremely powerful line emission in this object; in the $K$ band, 
the $Br\gamma$ line alone contributes between 6 and 7\% to the observed emission,
and the $V$ correction, including lines, is $-0.5$\,mag (Thuan et al. 1997).

To better assess the effects of these factors on the model predictions,
we have constructed $Z_\odot$/50 synthetic spectral energy distributions (SEDs)
which include stellar and ionized gaseous emission in proportions based on
our optical spectra. 
We have used the stellar SEDs calculated by 
Schaerer (2000, private communication) for a heavy element mass fraction of 
$Z$ = $Z_\odot$/50 and ages in the range of $t$ = 2 -- 100 Myr.
These SEDs are based on the same input physics as those for higher 
metallicities discussed by Schaerer \& Vacca (1998). 
The {\it observed} gaseous 
spectral energy distribution is then added to the calculated stellar spectral 
energy distribution, its contribution being determined by
the ratio of the observed equivalent width of the H$\beta$ emission line
to the one expected for pure gaseous emission. To calculate the gaseous 
continuum spectral energy distribution, the observed H$\beta$ flux and the 
electron temperature have been derived from the optical spectrum
(Izotov et al. 1997). 
The contribution of bound-free, free-free and 
two-photon continuum emission has been taken into account
for the spectral range from 0 to 5 $\mu$m (Aller 1984; Ferland 1980).
Emission lines are superposed on the gaseous continuum SED with 
intensities derived from spectra in the spectral range 
$\lambda$3700 -- 7500 \AA. Outside this range, the intensities
of emission lines (mainly hydrogen lines) have been calculated
from the extinction-corrected intensity of H$\beta$.
The transmission curves in the Johnson-Cousin-Glass photometric system
for the $U,B,V,R,I$ filters are taken from Bessell (1990) and for the 
$J,H,K$ filters from Bessell \& Brett (1988). The zeropoints are from 
Bessell, Castelli \& Plez (1998). 
The temporal evolution of the stellar population is shown in Fig.
\ref{vmijvk} as a dotted line; the dotted lines connecting the stellar
tracks to the filled circles show the gas contribution
and the filled circles the colors expected for gas$+$stars at
metallicity of $Z_\odot/50$ and 3~Myr of age, and
at 5~Myr (with  a larger extinction to simulate SSC~4+5).
Table~\ref{yuri} gives the colors predicted by this model for 
a 3~Myr $Z/Z_\odot\,=\,1/50$ stellar population together with the 
observed gas (continuum$+$line emission) fractions. 
The gas fraction in $K$ predicted by this model is around 70\%,
15\% or so larger than that observed ($\sim$\,50\%+6\%;
however, a contribution in $K$ by hot dust could explain the
discrepancy (see Sect.5.1).

\begin{table}
\begin{center}
\caption[]{Synthetic model predictions for SBS~0335-052 at
3~Myr and $Z/Z_\odot\,=\,$1/50.\label{yuri} }
\begin{tabular}{cccccc}
\hline
Process   & $V-I$ & $V-K$ & $J-K$ & $J-H$ & $H-K$ \\
\hline
Vega      & -0.009  & -0.008 & -0.002 & -0.003 &  0.001 \\
Stars$^a$ & -0.156  & -0.471 & -0.102 & -0.045 & -0.057 \\
Gas$^b$   &  0.576  & 2.097  & 1.034  &  0.406 &  0.628 \\
Gas$^c$   & -0.821  &  0.720 & 0.787  & 0.165  & 0.622 \\
Total$^d$ & -0.567  & 0.467  & 0.618  & 0.114  & 0.504 \\
\hline
\end{tabular}
\end{center}

\noindent
$^a$\ At $Z/Z_\odot\,=\,$1/50. 

\noindent
$^b$\ At 20000~K from Aller (1984) and Ferland (1980),
and with observed gas fraction and line emission,
but with continuum only.

\noindent
$^c$\ At 20000~K from Aller (1984) and Ferland (1980),
lines $+$ continuum.

\noindent
$^d$\ Stars $+$ gas (lines$+$continuum).

\end{table}

It is evident from Fig. \ref{vmijvk} that our model predicts the
colors of SSCs 1--2 and SSCs 4--5 better than the SB99 models.
The most likely explanation are the differences in the models:
most importantly the recombination line emission in the NIR,
but also the low metallicity and the high electron temperature of the gas,
all significantly alter the expected colors.
The global (E) and extended (Ex) emission in SBS~0335-052
are instead 0.3~mag redder in $J-K$ and $V-K$
than our model for the observed
$V-I\,\approx\,0$ (Thuan et al. 1997).
Part of this red color is almost certainly due to the surface brightness
limit in $K$; even with 55~minutes of on-source integration, SBS~0335-052
is so blue in the northwest, that we are not able to accurately measure
the colors\footnote{We find
a $V-K$ here of roughly 2, which is equivalent to our color surface brightness limit 
for a $K$ limit of 21.4~mag/arcsec$^2$, and $V$ of 23.4~mag/arcsec$^2$; it is
a strict upper limit.}
in those low surface brightness regions ($V\,\sim\,23.4$~mag/arcsec$^2$).
The $V-K$ in a higher brightness area ($V\,\sim\,22$~mag/arcsec$^2$), 
slightly closer to SSC~4+5, gives the $V-K$ of $\sim$1.3 shown in Fig. \ref{vmijvk}.
Such a $V-K$ color, for the measured $V-I$ of around zero, corresponds to a rather
old age in the SB99 models; however, there is probably hot dust
emission in the $K$-band (see Sect.5.1).

\subsection{Near-infrared colors}
\begin{figure}
\psfig{figure=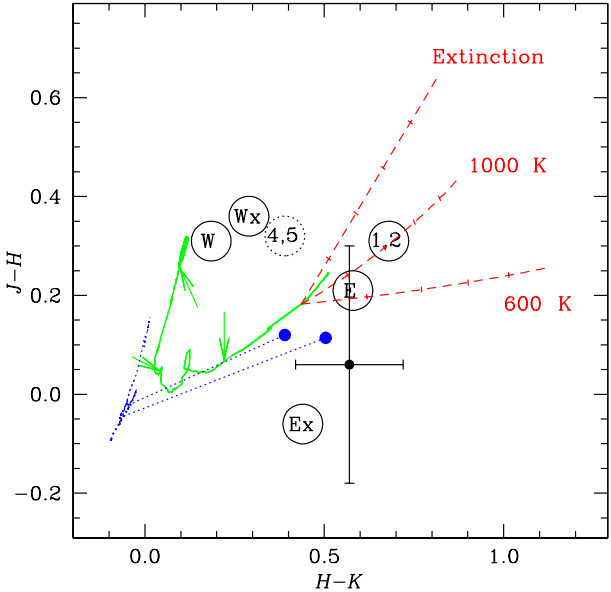,width=9cm,height=9cm,angle=0}
\caption{Near-infrared color-color diagram: $J-H$ vs. $H-K$.
Symbols as in Fig.\ref{vmijvk}, but with
``W'' and ``Wx'' for West global and extended.
The solid circle with the error bars for the NW extended region 
is derived from the color image. Colors are not corrected for gas emission. 
Our model and SB99 as in Fig. \ref{vmijvk}. 
Included also are (shown as dashed lines) dust extinction, 
and hot dust with emissivity$\,\propto\,1/\lambda$) at two temperatures.
The extinction curve tickmarks correspond to $A_V\,=\,$1 to 5 mags.
The tickmarks for the dust emission
denote the fraction of the total $K$-band due to dust,  
assuming a set of basis colors (in this case the 3~Myr colors of SB99);
the last tickmark is at equal contributions.
}
\label{jmhhmk}
\end{figure}

Fig.\ref{jmhhmk} shows the NIR colors of the various components of 
SBS~0335-052 and SBS~0335-052W, and illustrates how unusual the observed
colors are.
The sources shown in the figure
include SSCs 1+2, 4+5 (but this last with high uncertainty);
the global (E) and mean outer radial profile (Ex) of the eastern source;
analogous quantities for the western source (W, Wx, respectively);
and, finally, the diffuse region to the NW of 4+5 (as solid circle with
error bars).
Also shown in the figure (as solid lines)
are the SB99 $Z_\odot/20$ instantaneous models,
as well as mixing curves for various processes (using the 3~Myr SB99 colors
as the starting point).
Fig.\ref{jmhhmk} also shows (as dotted lines) our $Z_\odot/50$ stellar models,
together with the 3 and 5~Myr stellar population $+$ gas as described in the
previous section.
As before, our models with line emission
and at lower metallicity than SB99 seem to better predict
the colors of SBS~0335-052.

The NIR colors of SBS~0335-052 (1+2, E, Ex, but 4+5 uncertain)
shown in Fig. \ref{jmhhmk} are not well predicted by either set of models.
The colors of the global emission (``E") are close to those
predicted for a very young starburst (age $<\,4$~Myr), 
and are probably dominated by SSCs 1+2.
The extended component of the eastern source, shown in Fig. \ref{jmhhmk}
as ``Ex" and by the solid circle with error bars (the NW diffuse emission),
is not well modeled by the SB99 models, but at least the NW region
is close to our model predictions.
In any case, the $H-K$ colors of SSCs 1--2, E, and the NW region
are redder than either of the models.
A possible explanation for this discrepancy, suggested by the
dust mixing curves, is the presence of hot dust (see Sect.5.1).
Although the uncertainties are large, 
taken at face value, the NIR colors of the E component appear to
be those of very young (3~Myr) stars and a large gas fraction, 
together with a 20--30\% contribution of hot dust added.

On the other hand, the western source colors are much redder in $J-H$ and
bluer in $H-K$ than those of the eastern one.
The errors on the global colors are small ($\sim\,0.07$), so that this should
be a reliable statement.
For the same metallicity, these color changes correspond to an older age:
both the global and asymptotic profile colors appear to derive from
stellar populations of 10--20~Myr of age, since 
the points are near, but not on, the models of SB99 at this age.
However,
the $H\beta$ equivalent width, a good measure of starburst age,
also suggests that the W source is older than the E one
(Lipovetsky et al. 1999), but with an age not older than 5--7~Myr.
Indeed, our models reproduce approximately the NIR colors of
the center (not shown in Fig. \ref{jmhhmk}) of SBS~0335-052W 
with a single-age 6~Myr stellar population 
(plus gaseous emission): $J-H\,=\,0.09$, $H-K\,=\,0.34$
(c.f., Table \ref{tbl:phot_all} with $J-H\,=\,0.19$ and $H-K\,=\,0.27$
in $\phi\,=\,$4\arcsec).

\section{Stellar populations, star formation and dust in SBS~0335-052}

From the $V-K$ and $J-K$ colors in Fig. \ref{vmijvk}, we conclude 
that the SSCs 1+2 and 4+5 are very similar in age, roughly 3--4~Myr
according to our models and SB99.
Such a young age is consistent also with the spectral information:
we have detected no NIR absorption lines which would be typical of red
supergiants which onset at roughly 10~Myr (Maeder \& Meynet 1988). 

The extremely high EW of $Br\gamma$ ($>$~200~\AA ) also
suggests a very young age; indeed, the SB99 models predict an age of
between 3 and 4 Myr for such a high $Br\gamma$ EW
(instantaneous burst at $Z_\odot/20$ metallicity).
To better quantify this point we used the $Br\gamma$ flux and $K$ continuum
corrected for extinction and nebular emission to constrain the starburst model
of Rieke et al. (1993). We used a solar neighborhood IMF (n. 3) and a
burst duration of 1 and 5 Myr. Unfortunately the model is only available
for $Z_\odot$ metallicity, and the drawbacks of its application to low-metallicity
environments are discussed by Vanzi et al. (1996). The results are summarized 
in Table~\ref{george}, in which we have
derived total mass, bolometric luminosity, and age of
the burst for different regions of the galaxy. 
Indeed, as with the two other models, the stellar population age is between
3 and 4.5~Myr.
It also appears not be a particularly massive burst as total masses are
around $10^7\,M_\odot$.
From the values of Table ~\ref{george},
we can infer a star formation rate between 2 and 10 $M_{\odot}$/yr
for the whole galaxy. A higher correction to the $K$ continuum, due for
instance to dust, would reduce this estimate (see Sect.5.1).

\begin{table*}
\begin{center}
\caption[]{Parameters used to constrain the Rieke93 model and quantities 
derived. Results are for a burst of duration 1 and 5 Myr.
The Mass is espressed in $10^6~M_{\odot}$, the bolometric luminosity 
in $10^9~L_{\odot}$, the age in $10^6$ yr.} 
\label{george}
\begin{tabular}{ccccccccc}
\hline
region   & Log(UV) & K-40\% & M(1) & L(1) & age (1) & M(5) & L(5) & age(5) \\
\hline
Slit-LR  & 52.71   & 17.57  & 6.8  & 3.1  &  3.0    & 6.8  & 2.1  &  4.0 \\
Slit-SE  & 52.49   & 17.80  & 4.9  & 2.2  &  3.0    & 4.1  & 1.3  &  4.5 \\
Slit-NW  & 52.10   & 18.87  & 1.9  & 0.8  &  3.0    & 1.7  & 0.5  &  4.5 \\
Slit-Tot & 52.90   & 17.03  & 11.0 & 5.0  &  3.0    & 10.4 & 3.3  &  4.5 \\
\hline
\end{tabular}
\end{center}
\end{table*}


The $V-K$ and $J-K$ colors of the region extended to the northwest 
beyond SSCs 4+5 in the eastern source (see Fig. \ref{vmijvk})
seem to imply a young age similar to that estimated for the SSCs, 
but with an excess in $K$. 
Although such an excess could be imputed to our relatively low
sensitivity in $K$ to the blue colors observed in SBS~0335-052,
we find a $V-K\,\sim\,$1.3, even in relatively high surface brightness regions
where our deep UKIRT $K$ image is sufficiently sensitive.
The $H-K$ colors of the extended region (NW), shown in Fig. \ref{jmhhmk} are
very red, and $J-H$ is blue; the red $H-K$ is 
an indication that there may be hot dust also in the extended region.

We can estimate what fraction of an evolved stellar population
could be present in SBS~0335-052, given the observed NIR colors.
The NIR is the best spectral region to perform this estimate, since
it is more sensitive than the optical to evolved stars.
First, we find absolutely no color evidence for an evolved stellar population
(age greater than 10\,Myr) in any of the regions
examined, including the diffuse low-surface-brightness component.
Hence, the fraction of ``hidden" evolved stars that could be present
is essentially dictated by the uncertainties on our NIR colors.
In the SSCs and the global colors, 
these are nominally around 0.07~mag, since each magnitude has an
uncertainty of roughly 0.05~mag or better, including calibration errors.
Therefore, by adding the colors of an evolved ($>$~1\,Gyr) to the 3~Myr
SB99 color, we can estimate 
what fraction of an evolved stellar population would be responsible for
a color change in $J-H$ and $H-K$ that is equivalent to our measurement uncertainty. 
It turns out that 
$J-H$ changes by around 0.07~mag for a 16--18\% contribution of an evolved population
to the 3~Myr SB99 color; 
the concomitant $H-K$ color change is smaller (0.03~mag).
Because we are observing the {\it total} colors, the fraction becomes
$0.18/(0.18+1)$ or 15\%; that is to say we cannot exclude
a 15\% or smaller contribution to the NIR colors (of the SSCs or globally)
that derives from an evolved stellar population. 
This upper limit is unchanged if we add evolved NIR colors to a 
3~Myr solar-abundance stellar population (SB99), rather than to a metal-poor one.

We are also interested in the fraction of evolved stars that could be
present in the diffuse region that extends to the northwest.
As shown in Fig. \ref{jmhhmk}, the uncertainties there
are larger than those in the SSCs.
If we take an error of 0.25~mag in $J-H$, we derive a ratio of 0.9
between evolved and young stellar population colors.
This implies that 47\% of the extended emission could derive from an
evolved population, given the $J-H$ uncertainties in the extended region.
However, the $J-H$ colors are not the whole story.
We find an $H-K$ color of $>\,0.5$ in the outer regions, whether we use the
integrated profiles or the photometry of the color images to measure it.
Such a red color is not reproducible by adding an evolved stellar population
to the 3~Myr NIR colors;
indeed, the $H-K\sim0.2$ color of an evolved stellar population is {\it bluer} 
than that of stars$+$gas at 3~Myr. 
It is difficult to imagine a scenario involving a significant
evolved stellar population in which the colors are
conveniently uncertain in $J-H$, but provide such a red $H-K$.

\subsection{Hot dust in SBS~0335-052} 

To investigate the hot dust that we have tentatively invoked
to explain the red $H-K$ colors,
we have combined the mid-infrared ISO spectrum of SBS~0335-052
(Thuan et al. 1999) with the $K$-band emission after accounting
for the observed gas fraction.
After subtracting the gas contribution,
we have assumed that half of the remaining $K$ band flux is due to
hot dust.
The combined SED has been fitted with two modified blackbodies at different
temperatures and a foreground dust screen.
Dust temperatures and amplitudes, dust emissivity index, and visual extinction
$A_V$ were left as free parameters, and
we have adopted the extinction curve given in Draine (1989).

\begin{figure}
\psfig{figure=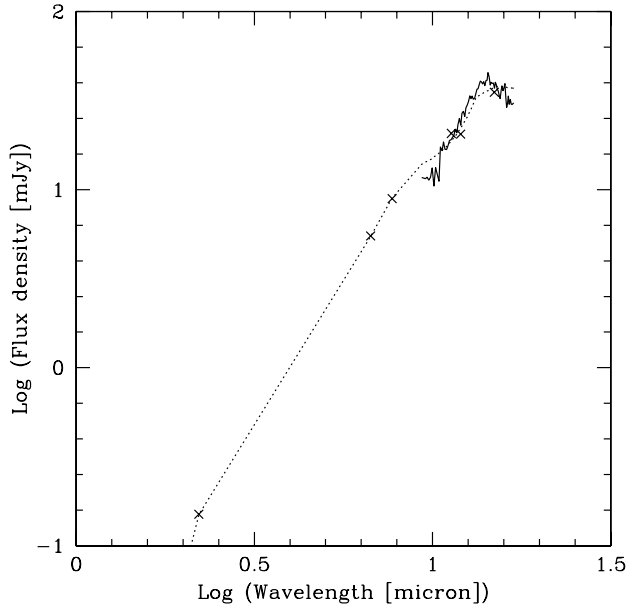,width=9cm,height=9cm}
\caption{Mid-infrared SED (Thuan et al. 1999) with the composite
fit described in the text shown as a dotted line.
The ISO CVF spectrum is shown as a solid line, and the broadband
photometric points given in Thuan et al. (1999) and in this
paper ($K$) as $\times$.}
\label{dust}
\end{figure}

The resulting fit is shown in Fig. \ref{dust} as a dotted line.
The warm and hot components have temperatures of 226~K and 668~K,
the cooler temperature being similar to 260~K found by Thuan et
al. (1999).
The fitted emissivity is around $\nu^{0.5}$ (instead of the $\nu^{1.5}$ used
by Thuan et al. 1999), and the amplitude of the hotter component
is 9\% that of the cooler one at 10\,$\mu$m.
At shorter wavelengths, the ratio of the hotter to the cooler component
increases, being 31\% at 7.7\,$\mu$m and 68\% at 6.7\,$\mu$m.
The fitted extinction $A_V$ is around 12~mag, but more uncertain
($\pm\,$1~mag) than the temperatures (uncertainties $\pm\,$10~K).
In any case, without a significant extinction, the SED is very poorly
fit as it is too narrow around its peak at 10~$\mu$m.
With two dust components, we find a slightly lower temperature
(for the cooler dust) and lower extinction than the $A_V$ of 21~mag
found by Thuan et al. (1999).
Nevertheless,
both models show that the mid-infrared extinction is high, in
variance with what we infer from the optical/NIR recombination lines.
The most likely explanation is the clumpy nature of the interstellar medium,
since we do not detect peaks of emission at 2\,$\mu$m corresponding
to hidden SSCs.
There may be a relatively high quantity of dust globally, but
not locally in the vicinity of the SSCs we detect.

The fit of the mid-infrared spectrum around 10~$\mu$m is poor in detail 
because the Draine extinction curve overpredicts silicate absorption
there (Lutz et al. 1996), and indeed fits of the SED in SBS~0335-052
with the Galactic Center curve give better results (Thuan et al. 1999).
However, it is clear that a hot-dust component can fit the
shorter-wavelength points very well, and consistently with
the $K$ emission after gas subtraction.
Indeed, such dust seems to be necessary to explain the red $H-K$
colors of SBS~0335-052.

\section{Conclusions}
We can summarize the results of our work as follows:

\begin{enumerate}
\item The near-infrared spectrum of SBS~0335-052 is representative of an extreme
and young starburst galaxy. 
The $Br\gamma$ equivalent width indicates a very young age, and is one of the
largest ever observed in an extragalactic object. 
The He I recombination lines support the presence of young massive stars.
Recombination and $H_2$ lines are strong, [FeII] and stellar
absorption bands are absent, which also implies a strong UV field 
and a young stellar population;
there is no spectroscopic evidence for stars older than 5--6~Myr.
Modeling the $Br\gamma$ and $K$ continuum leads
to a burst age$\leq$5 Myr and a star formation rate that can be as high as 10
$M_{\odot}$/yr. 
\item The spatial profile of the $K$ continuum is shifted by about 60 pc
northwest of the emission line peak and it is much closer to it than the
optical peak at a distance of 200 pc. This is explained with the dominant
nebular contribution to $K$. All optical and NIR emission lines coincide
and have consistent spatial profiles.
\item The near-infrared photometry of SBS~0335-052 is highly contaminated
by nebular emission. The optical and NIR colors are unusual, and after 
correction for the gas contribution can be only understood as due to 
a stellar population not older than 4 Myr, together with hot dust at
670~K.
\item
On the basis of the NIR colors, the stellar population in
SBS~0335-052W appears to be older
than that in SBS~0335-052; the colors are redder in $J-H$ and bluer in $H-K$ and
correspond to an approximate age of 10--20~Myr according to SB99.
However, optical recombination line equivalent widths suggest that
the stellar population cannot be older than 10~Myr, and our models
suggest that the age is more likely around 5--7~Myr. 
\item The gas-corrected spectrum from 2 to 17\,$\mu$m is well fit by
two thermal components (hot dust at 670~K, and cooler dust at 225~K), together
with $A_V\,=\,12$\,mag of visual extinction in a foreground screen.
\item We do not find evidence at 2\,$\mu$m for optically hidden star 
formation. 
\item Judging from the uncertainties on the NIR photometry,
the possible contribution from an evolved stellar population in
SBS~0335-052 cannot exceed 15\%.
\end{enumerate}

SBS~0335-052 appears to be an unusual object, as there is no evidence,
even in the near-infrared, for an evolved stellar population deriving
from an earlier burst.
This is in contrast to results for the lowest-metallicity galaxy
known, I~Zw~18, in which Aloisi et al. (1999) and \"Ostlin (2000) claimed 
to detect a 1--4 Gyr underlying stellar population. 
If the star-formation rates derived from our modeling are representative,
then stars are forming in SBS~0335-052 extremely rapidly, but
how such a rate is connected to the formation of SSCs is an open question.
It could be that under certain conditions, star formation, 
galaxy gas consumption, and evolution proceed very quickly, while in
others, evolution is slower.
Future work will be aimed at better quantifying what these conditions
might be, and how they relate to metallicity.

\begin{acknowledgements}
      We thank Gian-Paolo Tozzi who generously shared with
      us part of his observing time at the NTT during the run of August 
      1998. 
      TXT thanks the partial financial support of NSF grant AST-9616863 and
      the hospitality of the Observatoire de Paris-Meudon and the
      Institut d'Astrophysique de Paris during his sabbatical year.
      We are also grateful to George Rieke who made available to
      us the results of his starburst model, to Daniel Schaerer who
      gave us his low-metallicity stellar tracks and to Leonardo Testi
      for useful comments on the manuscript.
\end{acknowledgements}

\end{document}